\documentclass{aa} 

\usepackage{marvosym,epsfig,amssymb}
\usepackage{txfonts}
\usepackage[figuresright]{rotating}
\usepackage{natbib}
\bibpunct{(}{)}{;}{a}{}{,} 

\def \sol{\ifmmode _{\mathord\odot} \else $_{\mathord\odot}$\fi}
\def\lea{\mathrel{\raise .4ex\hbox{\rlap{$<$}\lower 1.2ex\hbox{$\sim$}}}}
\def\gea{\mathrel{\raise .4ex\hbox{\rlap{$>$}\lower 1.2ex\hbox{$\sim$}}}}

\begin{document}

\title{Star-forming QSO host galaxies}

\author{P.D.~Barthel}

\institute{Kapteyn Astronomical Institute, University of Groningen,
P.~O.~Box 800, 9700 AV Groningen, The Netherlands}

\offprints{P.D.~Barthel, \email{pdb@astro.rug.nl}}

\titlerunning{QSO hosts}
\authorrunning{P.D.~Barthel}

\date{Received date / Accepted date}

\abstract{The recent finding of substantial masses of cold molecular gas
as well as young stellar populations in the host galaxies of
intermediate luminosity quasars is at odds with results of Hubble Space
Telescope imaging studies since the latter appear to yield mature,
quiescent early type hosts.}{The characterization as `quiescent' for the
host galaxies of QSOs is being addressed.}{The radio and far-infrared
properties of both the HST sample and a larger comparison sample of
uv-excess selected radio-quiet QSOs are being analyzed.}{Consistency is
found with substantial recent or ongoing star-formation activity.}{}


\maketitle

\section{Introduction}

Quasi-stellar objects, or QSOs, represent episodes of extreme nuclear
luminosity in galaxies, over a wide wavelength range.  QSOs display a
considerable dispersion in range-loudness, which is commonly expressed
(Kellermann et al.  1989) as ratio $R$ of cm-radio over blue (4400\AA)
flux density.  Radio-quiet QSOs typically have $R \lesssim 1$, while
radio-loud QSOs are characterized with $R$ in excess of 10.  QSO host
galaxies -- both of the radio-loud and the radio-quiet subgroup -- have
been targets of studies for many years.  Early investigations (e.g.,
Hutchings et al.  1984) revealed luminous hosts, frequently with close
companions, and sometimes displaying spectroscopic features of young
stars.  Systematic studies of QSO host galaxies became possible with the
refurbished Hubble Space Telescope (HST): host galaxies of different
nature were identified (Bahcall et al.  1997).  Using bulge--black hole
mass correlations, these studies permitted assessment of the QSO
accretion rates, and rates up to 20 per cent of the Eddington values
were inferred (McLeod et al.  1999).  Ground-based near-infrared studies
of samples of both radio-loud and radio-quiet hosts as well as powerful
(P$_{\rm radio} > 10^{27}$ W/Hz) radio galaxy hosts indicated (Taylor et
al.  1996) the predominance of very luminous spheroidal hosts,
particularly in the two radio-loud AGN populations.  These results were
confirmed by imaging studies from space, using HST.  In the final paper
(Dunlop et al.  2003) in a series dealing with WFPC2 studies of the host
galaxies of matched samples of radio-loud and radio-quiet QSOs as well
as radio galaxies, it was concluded from 2-D model fitting procedures
that "...  for nuclear luminosities brighter than $V=23.5$ the hosts are
virtually all massive ellipticals with properties indistinguishable from
those of quiescent, evolved low-redshift ellipticals of comparable
mass."

Powerful radio-galaxies are long known to be hosted by massive
ellipticals, which are however sometimes of disturbed morphology (e.g.,
Heckman et al.  1986).  While the HST imaging results (Dunlop et al. 
2003) seem consistent with AGN unification models (involving obscuring
dusty circumnuclear tori -- e.g., Barthel 1989, Urry \& Padovani 1993)
it must be stressed that also the presence of extended dust, cold gas,
and young stars in many radio-loud AGN has been reported (Hes et al. 
1995, Martel et al.  1999, Tadhunter et al.  2005, Van Gorkom et al. 
1989), which in turn is difficult to reconcile with the picture of
quiescent ellipticals.  Similar contrasting views exist on the hosts of
radio-quiet QSOs, i.e., QSOs displaying radio-optical flux ratio's of
order unity or smaller.  Whereas -- as mentioned above -- 2-D model
fitting of the red starlight distributions generally seems to yield
dominant spheroids, the properties of the interstellar media of several
radio-quiet QSO hosts do not support the quiescent massive elliptical
classification.  The recent finding (Evans et al.  2001, Scoville et al. 
2003) of substantial masses of cold molecular gas, as well as several
cases (Canalizo \& Stockton 2001) of young circumnuclear stars in
radio-quiet QSO host galaxies is at odds with conclusions of the imaging
studies.  Some radio-quiet QSOs must be (Clements 2000) post-starburst
systems, of which QSO UN J1025--440 provides a spectacular example
(Brotherton et al.  1999).  An extensive radio-optical study (Miller et
al.  1993) using 89 low redshift QSOs, indicated that roughly half of
the weak radio emission in the radio-quiet subgroup might draw from host
star-formation.  A study of the integrated radio and far-infrared
emission in various classes of active objects (Colina \& Perez-Olea
1995) confirmed that suspicion.  The HST results (Dunlop et al.  2003)
are also inconsistent with modeling studies (Rowan-Robinson 1995) of
infrared spectral energy distributions (SEDs), concluding "...  an AGN
is inevitably accompanied by a star-burst ..."

We therefore examine in detail radio imaging and far-infrared
photometric data for the radio-quiet subset of the HST QSO sample of
Dunlop et al.  (2003), and demonstrate from comparison with similar data
for the uv-excess selected PG QSOs that the radio and far-infrared
properties of both samples are in fact consistent with the presence of
gas-rich star-forming interstellar media in the AGN hosts.  So while we
acknowledge the presence of a massive spheroid as a necessary condition
for the nuclear activity to occur, we conclude that the circumnuclear
regions in the host galaxies of intermediate luminosity QSOs must
contain young stars, supernova remnants, and starburst heated dust. 
That conclusion is also in line with results of recent SDSS studies
(Kaufmann et al.  2003), suggesting that many AGN are post-starburst
systems, and recent ISO studies (Haas et al.  2003). 

\section{Sample and analysis}
\label{sec:analysis}

The HST/Dunlop et al.  radio-quiet (i.e., radio-weak, not radio-silent)
QSO sample (Dunlop et al.  2003) consists of thirteen objects, in the
redshift range $0.1<z<0.25$.  Of these thirteen, QSO 1635+119 is
associated with an extended ($\sim$1~arcmin) radio source of luminosity
P(5\,GHz) = $5\times10^{24}$ W/Hz; combining this with its $R$(5\,GHz)
value of 67, we contend that this object should be formally classified
as a radio-loud AGN (its optical magnitude is too faint for a QSO). 
Note that radio-emission in radio-quiet QSOs is generally weak (in
absolute and relative terms), and confined to the nuclear and/or
circumnuclear regions of the QSO host galaxy.  The thirteen HST QSOs
will be compared with Palomar-Green (PG -- Schmidt \& Green 1983) QSOs
in the same redshift range, $0.1<z<0.25$.  The latter is a well known
sample of QSOs selected through their ultraviolet-excess, and has
well-documented properties (Sanders et al.  1989): thirty-one
radio-quiet PG QSOs are found in the relevant redshift range.  Both
samples are tabulated in the Table (note: six QSOs in common).  Besides
name, redshift, and absolute optical magnitude, we list flux densities
at 60$\mu$m and 5\,GHz, inferred luminosities at 60$\mu$m and 5\,GHz,
radio core fractions $f_c$ at 5\,GHz, u-parameters (logarithmic ratio's
of the 60$\mu$m and 5\,GHz flux densities), for the integrated (u) and
the extended, radio-core subtracted (u$'$) emission, and relevant
references.  Core fractions $f_c$ were obtained by comparing the 5\,GHz
flux densities obtained at low resolution (typically 18~arcsec --
Kellermann et al.  1989) with those at high resolution ($\lesssim$
0.5~arcsec).  In several cases IRAS ADDSCAN routines were used to
re-assess the 60$\mu$m properties, yielding new detections as well as
improved upper limits.  Having $-25 \lea {\rm M}_{abs} \lea -23$, the
objects comprise a representative sample of intermediate luminosity
radio-quiet QSOs. 

\section{Results}
\label{sec:results}

The 60$\mu$m far-infrared luminosities, for the HST and the PG samples,
run from 10.5 to 12.5 (in log L\sol).  The former has five upper limits,
the latter eight.  The radio luminosities run from $10^{22}$ to
$10^{24}$ W/Hz, with one $10^{24.7}$ in the Dunlop set (the radio-loud
object).  Although the fractions of 60$\mu$m non-detections (5/13 vs. 
8/31) suggest otherwise, survival analysis (taking the upper limits into
account) cannot exclude that the two samples were drawn from one and the
same parent population.  More sensitive infrared photometry is needed to
re-address that issue.  The FIR and radio luminosities however, indicate
that the hosts cannot be quiescent early type galaxies.  The Dunlop and
PG QSO sample display 60$\mu$m luminosities which are orders of
magnitude in excess of (IRAS/ISO) luminosities for massive inactive
ellipticals, which are generally (Walsh et al.  1989) in the range
$10^7$ -- $10^9$ L\sol.  There would be an obvious explanation for this
fact, if all far-infrared and radio emission drew from the nuclear
activity, but this appears not to be the case as we will demonstrate in
the following.

\begin{figure}
\scalebox{0.48}{\includegraphics{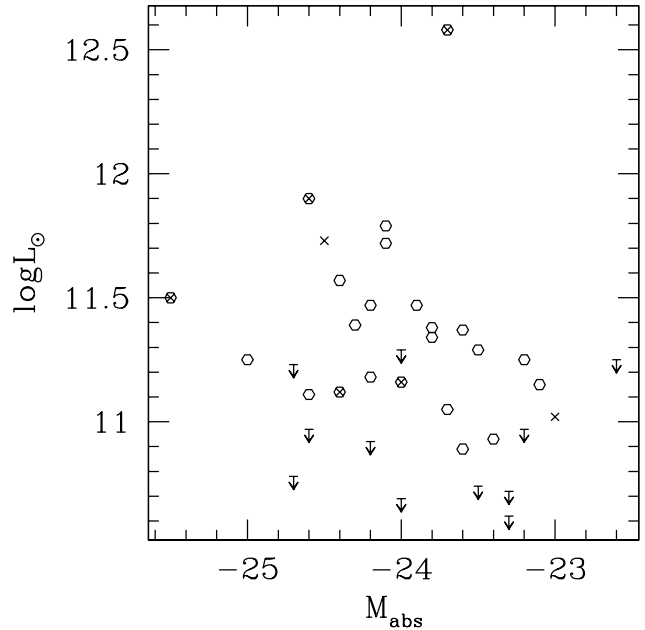}}
\caption{60$\mu$m FIR luminosities (expressed in log L\sol) versus
absolute (nuclear) optical magnitudes, for the QSOs from both samples
(with the exception of 0257+024, 1635+119 and 2344+184, whose optical
faintness formally disqualifies these object as QSOs).  The crosses mark
QSOs from the HST sample, the open hexagons mark QSOs from the
PG-sample.  For the sake of simplicity, the R-values for the HST sample
are combined with the B-values for the PG-sample (R-values used for the
common objects), since the (small) color transformations are uncertain
due to intrinsic spread in the optical continuum slopes and varying
emission line contributions.}
\label{FigLum}
\end{figure}

Fig.~\ref{FigLum} compares the optical nuclear luminosities and the
60$\mu$m luminosities, for the QSOs in both samples.  While the two
quantities are broadly correlated, a substantial scatter is seen. 
Fig.~\ref{FigLum} -- dealing with a small sample in a fairly narrow
redshift range -- confirms results from the Haas et al.  (2003) study of
a large sample of PG QSOs studied with ISO.  This study reports the
presence of high infrared luminosities implying large dust masses
ranging from $10^6$ to $10^{8.5}$ M\sol.  However, in contrast to the
near-IR (3--10$\mu$m) and the mid-IR (10--40$\mu$m) luminosity, being
fairly well correlated with the QSO (blue) luminosity, the far-IR
emission (40--150$\mu$m) is found to display a scatter of about a factor
10 in its correlation with that QSO luminosity.  This suggests that
FIR-emission surely draws from the active galactic nucleus but must in
addition have a significant contribution from stellar heating.  As Haas
et al.  (2003) stress, the large amounts of dust in QSOs would require
that the QSO phase was at least {\it preceeded} by a starburst.  We
investigate this further by addressing the radio continuum emission as
this offers an unobscured probe of host galaxy star-formation. 

As demonstrated earlier (e.g., Wilson 1988), radio-emission in active
galaxies can be separated into two components, one deriving from the
nuclear activity and one from host galaxy star-formation.  Subtracting
the AGN-driven component from the integrated radio emission, an improved
correlation with respect to the integrated long wavelength far-infrared
emission (which draws from star-formation) is being found.  The
u-parameter, as defined above, is generally used for this purpose.  This
parameter relates the 60$\mu$m FIR emission to the 20cm or 6cm radio
emission.  Star-forming objects are characterized (Condon \& Broderick
1988) with average (20cm) parameter u$_{20}=2.0$, and $\pm2\sigma$-range
1.6--2.4, i.e., 1.4\,GHz radio emission with strength between 0.4 and
2.5 per cent of the 60$\mu$m FIR emission (1 per cent in the mean). 
Using the standard radio spectral index $-0.7$ for optically thin
synchrotron emission, this translates to 6cm (5\,GHz) parameters in the
range 1.95 $<$ u$_6 <$ 2.75 (mean 2.35).  In fact, from radio imaging
and FIR SED decomposition it is well known that the FIR emission
combines an AGN- with a star-formation component.  Objects with an
appreciable active nucleus display surplus radio emission, hence
u-parameters $\lesssim 1$.  In Seyfert-2 galaxy NGC\,1068 for instance,
roughly 80\% of its 20cm radio emission draws from the 1.5~kpc triple
radio source, hence is generated by the AGN.  The remaining 20\% occurs
on larger scales, and finds its origin in host star-formation (Telesco
et al.  1984).  Prototypical ultraluminous starburst galaxy Arp\,220
(Sanders \& Mirabel 1996) has u$_6$ = log S(60$\mu$m)/S(5\,GHz) = log
(103.3Jy/0.204Jy) = 2.70 (hence negligible AGN component), star-forming
QSO Mkn\,231 (Carilli et al. 1998) has u$_6$ = log S(60$\mu$m)/S(5\,GHz)
= log (35.4Jy/0.419Jy) = 1.93 but u$'_6$ = log S(60$\mu$m)/S(5GHz,
extended) = log (35.4/0.173) = 2.31.  In words, roughly 40\% of the
5\,GHz radio emission in Mk\,231 (and most of the long wavelength FIR
emission) must draw from star-formation. 

We therefore use sensitive, high resolution radio imaging data
(references in Table column 12) to assess -- where possible -- the
radio-core fractions ($f_c$) and to transform the global, integrated
u-parameters into AGN-subtracted u$'$-parameters.  It must be noted that
we are dealing here with radio-quiet objects, i.e., objects without
large scale radio lobes or jets, and that weak AGN radio emission -- if
any -- is most likely concentrated in faint subarcsec scale radio
components (cores hereafter).  Inspection of the $f_c$ entries shows
several blanks in the case of objects lacking adequate radio imaging,
but on the other hand very few pure-core ($f_c = 1$) objects: diffuse
galactic scale radio emission must be occurring often.  Given that no
cases of $f_c > 1$ are seen, the $f_c$ values cannot be attributed to
core variability combined with different observing epochs: they must
reflect the presence of weak, diffuse, arcsec-scale (host galaxy scale)
radio emission which is resolved out at high angular resolution. 
Typical diffuse luminosities are of order $10^{22}$ -- $10^{23}$ W/Hz,
i.e., roughly one to ten times the diffuse radio emission in our Galaxy. 
Inspection of the Table moreover shows that the extended (or integrated)
radio emission frequently correlates with the integrated 60$\mu$m
emission according to the radio-FIR correlation: many cases of u$_6$ and
u$'_6$ in the range 2.0 -- 2.8 are seen.  The few cases of exception
(PG\,0923+201, PG\,0947+396, PG\,1114+445, PG\,1115+407, PG1435$-$067)
display suspiciously high FIR emission which is probably due to
contamination in the extended infrared beam.  A few cases of mildly
excessive radio emission (1.5 $<$ u$_6<$ 2) can be also seen. 

As examples we consider the QSOs PG\,0052+251, PHL\,909 (0054+144), and
PG\,1012+008.  Detailed HST images of the first two objects have been
presented (Bahcall et al.  1996), establishing a spiral host for the
former and an elliptical host for the latter.  PG\,0052+251 displays
several HII regions in the disk of its host, and the HST imagery (Dunlop
et al.  2003) indeed yielded a bulge/disk composite.  As for PHL\,909,
the 2D-analysis (McLure et al.  1999) improved on the earlier (Bahcall
et al.  1996) HST picture, excluding a disc host and establishing an $R
= -23.62$ elliptical host having an 8\,kpc half-light radius.  As seen
from the Table, both objects have significant diffuse host galaxy radio
emission, and their AGN-corrected u$'_6$-parameters fall nicely on the
FIR-radio correlation.  In particular, the luminous FIR emission of
PHL\,909 as well as the fact that 50\% of its 6cm radio emission must
originate in extra-nuclear regions are noteworthy.  While there cannot
be any doubt on a disk+bulge host for PG\,0051+251, also the host of
PHL\,909 cannot be a quiescent mature elliptical.  We recall that the
residual images (McLure et al.  1999, Dunlop et al.  2003) of PHL\,909
display faint excess emission towards the West; additional excess
emission may well be present under the QSO PSF.  PG\,1012+008 resides in
a host galaxy which was found (Bahcall et al.  1997) to be in the
process of interaction with a neighbouring galaxy.  Roughly one quarter
of the radio emission has a diffuse arcsec scale origin, scaling with
the 60$\mu$m emission cf. the radio-FIR correlation. 

In other words, the radio/FIR differences with quiescent ellipticals
cannot simply be attributed to the AGN.  A substantial fraction of the
radio and FIR emission in these intermediate luminosity QSOs must draw
from star-formation in their hosts.  While in some cases this
star-formation can be readily connected to residual emission in the HST
imagery (after subtraction of the spheroids), in other cases it must
connect to small residuals under the PSF, i.e., to star-formation within
the central 100 parsecs.  Given that circum-nuclear star-formation is
frequently observed in Seyfert galaxies (e.g., Gonz\'alez Delgado et al. 
2001), in connection with the fact that Seyfert galaxies are generally
considered to be the low-luminosity counterparts of QSOs, these results
are not unexpected.  We stress that we are not questioning the presence
of massive bulges in the QSO hosts (and the probable connection (Dunlop
et al.  2003) of the bulge/disk ratio to the QSO luminosity), but we
feel urged to point out the special nature of the host ISM.  Back of the
envelope calculations indicate SFR's of order 10\, M\sol~per~year.  It
should be noted however that more luminous QSOs (brighter than $-25$)
probably do not have scaled-up SFR's (Haas et al.  2003), unless at
early epochs (e.g., Cox et al.  2005).  Our results are nevertheless
consistent with the multicolour imaging obtained by Jahnke et al. 
(2004a) for low redshift QSOs, as well as for more distant AGN by Jahnke
et al.  (2004b) and S\'anchez et al.  (2004).  It is tempting to
interpret this and other evidence for circumnuclear star-formation
within the AGN accretion model of Miralda-Escud\'e \& Kollmeier (2005). 

Finally, the above mentioned typical QSO SFR, computed from FIR data,
is incompatible with the value obtained using the [OII]$\lambda$3727
emission line strength of these QSOs.  This fact -- pointed out by Ho
(2005) -- calls for some level of extinction towards the [OII] emission,
which is not unexpected given the large dust masses suggesting dusty
young star clumps as f.i. also inferred in the Antennae galaxies. 

\section{Summary}
\label{sec:summary}

We conclude that a QSO episode is not seldom accompanied or preceded by
an episode of star-formation in the gas-rich circumnuclear regions of
its host galaxy.  We anticipate that radio-loud QSOs will show the same
effects, but note that radio imaging separating AGN and star-formation
driven radio emission in these objects is not going to be simple. 
Decomposition of the FIR SEDS will be more promising in this respect:
results from currently as well as soon flying infrared space
observatories are eagerly awaited.  The evolving dust and stellar
content of QSO host galaxies will remain an issue of great importance
and interest (e.g., Barthel \& Sanders 2006). 

\begin{acknowledgements}

This research was initiated during a work visit to the Institute for
Astronomy at the University of Hawaii; the hospitality and support of
Drs.  R.-P.~Kudritzki and D.B.~Sanders at the IfA are gratefully
acknowledged. 

\end{acknowledgements}

\begin{table*}
\label{Fulltable}

\caption{The Table lists integrated flux densities, in mJy, at 60$\mu$m
FIR and 5\,GHz (6cm) radio, radio core fractions $f_c$ (fraction of
5\,GHz radio emission in subarcsec core), luminosities at 60$\mu$m (in
solar units) and spectral power at 5\,GHz (W/Hz).  Furthermore,
u$_6$-parameters, addressing conformity to the FIR-radio correlation for
the integrated (u) as well as the core-subtracted emission (u$'$). 
Parameter u$_6$ is defined as log S(60$\mu$m)/S(5\,GHz), whereas u$'_6$
= log S(60$\mu$m)/S(5\,GHz, extended); star-formation generally yields
u$_6$-parameters in the range 2.0 -- 2.8; radio-loud AGN (with strong
non-thermal core and lobe emission) have u$_6$ $\ll 2$, incl.  negative
values.  Values for the Hubble parameter and deceleration parameter of
50 and 0.5 respectively, in a $\Lambda=0$ universe have been used to
compute source intrinsic quantities.  The 13 QSOs from the HST sample
(Dunlop et al.  2003) are followed by the 31 PG (Schmidt \& Green 1983)
QSOs: the former have absolute (nuclear) R-band magnitudes listed, the
latter B-band, or V-band (PG\,0906+484).  References: BWE91 -- Becker et
al.  1991, D03 -- Dunlop et al.  2003, K89 -- Kellermann et al.  1989,
Ku98 -- Kukula et al.  1998, S89 -- Sanders et al.  1989, T96 -- Taylor
et al.  1996}

\begin{tabular}{llrrrrrrrrll}
\hline\hline
QSO & $z$ & M$_{abs}$ & S(60$\mu$) & S(5GHz) & log\,L(60) & log\,P(5) & $f_c$ & u$_6$ &
 u$'_6$ & ref(60$\mu$) & ref(5GHz) \\
\hline
PG\,0052+251 & 0.155 & $-$24.4 &   93 & 0.74   & 11.12   &  22.88  & 0.57 &  2.10  & 2.47  & S89     &  K89       \\
0054+144     & 0.171 & $-$24.5 &  307 & 1.5$^a$ & 11.73 & 23.28 & 0.5 & 2.31 & 2.61 & FSC$^b$ & NVSS$^c$, Ku98    \\
PG\,0157+001 & 0.164 & $-$23.7 & 2377 & 8.0    & 12.58   &  23.97  & 0.4$^d$ & 2.47& 2.69  & S89     &  K89       \\
0204+292     & 0.109 & $-$23.0 & 150 &$\lesssim$10 & 11.02 &$\lesssim$23.7 & & $\gtrsim$1.2 & & ADDSCAN$^e$ & T96 \\
0244+194     & 0.176 & $-$23.2 & $<$50 & 0.18  &$<$10.97 &  22.40  &      & $<$2.4 &       & ADDSCAN &     D03    \\
0257+024     & 0.115 & $-$19.7 & $<$50 &  3.4  &$<$10.59 &  23.29  &      & $<$1.2 &       & ADDSCAN &    Ku98    \\
PG\,0923+201 & 0.190 & $-$24.6 & 361  & 0.25   & 11.90   &  22.59  & 0.6  &   3.2  & 3.5   & FSC     & K89, D03   \\
PG\,0953+414 & 0.239 & $-$25.5 &  90  &  1.9   & 11.50   &  23.67  & 0.1  &  1.68  & 1.75  & ADDSCAN & K89, D03   \\
PG\,1012+008 & 0.185 & $-$24.0 &  70  &  1.0   & 11.16   &  23.17  & 0.74 &  1.85  & 2.43  & ADDSCAN &  K89       \\
1549+203     & 0.250 & $-$24.0 & $<$50 & $<$0.12&$<$11.29 &$<$22.5 &      &        &       & ADDSCAN &  D03       \\
1635+119     & 0.146 & $-$21.5 & $<$50 &  49   &$<$10.80 &  24.65  &  RL  & $<$0   &       & ADDSCAN &  BWE91     \\
2215$-$037   & 0.241 & $-$22.6 & $<$50 & 0.13  &$<$11.25 &  22.50  &      & $<$2.6 &       & ADDSCAN &    D03     \\
2344+184     & 0.138 & $-$20.3 & 100  & 0.19   & 11.05   &  22.20  &      &  2.72  &       & ADDSCAN &    D03     \\
             &       &         &        &      &         &         &      &        &       &         &            \\
PG\,0026+129 & 0.142 & $-$24.7 & $<$50 & 5.1   &$<$10.78 &  23.65  & 0.04 & $<$1.0 &$<$2.4 & ADDSCAN &     K89    \\
PG\,0052+251 & 0.155 & $-$24.4 &  93  & 0.74   &   11.12 &  22.88  & 0.57 & 2.10   & 2.47  & S89     &     K89    \\
PG\,0157+001 & 0.164 & $-$24.8 & 2377  & 8.0   &   12.58 &  23.97  & 0.4$^d$ & 2.47 & 2.69 & S89     &     K89    \\
PG\,0804+761 & 0.100 & $-$23.7 & 191  & 2.38   &   11.05 &  23.00  & 0.41 & 1.83   & 2.13  & S89     &     K89    \\
PG\,0838+770 & 0.131 & $-$23.2 & 174  & $<$0.4 &   11.25 &$<$22.5  &      &$>$2.6  &       & S89   & NED$^f$, K89 \\
PG\,0906+484 & 0.118 & $-$23.1 & 172  & $<$0.5 &   11.15 &$<$22.5  &      &$>$2.5  &       & S89     &     NED    \\
PG\,0923+201 & 0.190 & $-$24.2 & 361  & 0.25   &   11.90 &  22.59  & 0.6  &  3.2   & 3.5   & FSC     &  K89, D03  \\
PG\,0947+396 & 0.206 & $-$24.1 & 201  & 0.31   &   11.72 &  22.75  & 0.77 & 2.81   & 3.45  & S89     &    K89     \\
PG\,0953+414 & 0.239 & $-$25.7 &  90  & 1.9    &   11.50 &  23.67  & 0.1  & 1.68   & 1.75  & ADDSCAN &  K89, D03  \\
PG\,1001+054 & 0.161 & $-$23.8 & 140  & 0.80   &   11.34 &  22.95  & 0.3  & 2.24   & 2.40  & NED     &     K89    \\
PG\,1012+008 & 0.185 & $-$24.3 &  70  & 1.00   &   11.16 &  23.17  & 0.74 & 1.85   & 2.43  & ADDSCAN &     K89    \\
PG\,1048+342 & 0.167 & $-$24.2 & $<$50 & 0.01  &$<$10.92 &  22.40  &      & $<$3.7 &       & ADDSCAN &     K89    \\
PG\,1114+445 & 0.144 & $-$23.6 & 191  & 0.22   &   11.37 &  22.28  & 0.91 & 2.94   & 3.98  & S89     &     K89    \\
PG\,1115+407 & 0.154 & $-$23.8 & 170  & 0.30   &   11.38 &  22.4   & 0.8  & 2.75   & 3.45  & ADDSCAN &     K89    \\
PG\,1121+422 & 0.234 & $-$24.7 & $<$50 & $<$0.1 &$<$11.23 &$<$22.4 &      &        &       & ADDSCAN &     K89    \\
PG\,1151+117 & 0.176 & $-$24.6 & $<$50 &  0    &$<$10.97 &$<$21.1  &      &        &       & ADDSCAN &     K89    \\
PG\,1202+281 & 0.165 & $-$25.0 & 110  & 0.83   &   11.25 &  22.9   & 0.78 & 2.12   & 2.78  & S89     &     K89    \\
PG\,1307+085 & 0.155 & $-$24.6 &  90  & 0.35   &   11.11 &  22.64  &      & 2.41   &       & ADDSCAN &     K89    \\
PG\,1322+659 & 0.168 & $-$24.2 &  90  & 0.20   &   11.18 &  22.38  &      & 2.65   &       & ADDSCAN &     K89    \\
PG\,1352+183 & 0.158 & $-$24.2 & 197  & 0.25   &   11.47 &  22.43  &      & 2.90   &       & NED     &     K89    \\
PG\,1402+261 & 0.164 & $-$24.4 & 229  & 0.62   &   11.57 &  22.8   & 0.4  & 2.57   & 2.79  & S89     &     K89    \\
PG\,1415+451 & 0.114 & $-$23.4 & 112  & 0.40   &   10.93 &  23.3   & 0.6  & 2.45   & 2.85  & S89     &     K89    \\
PG\,1416$-$129 & 0.129 & $-$24.0 & $<$50 & 3.60 &$<$10.69 & 23.4   & 0.22 &$<$1.14 &$<$1.25& ADDSCAN &     K89    \\
PG\,1427+480 & 0.221 & $-$24.3 &  82  & 0.02   &   11.39 &  21.6   &      & 3.6    &       & ADDSCAN &     K89    \\
PG\,1435$-$067 & 0.129 & $-$23.9 & 304  & 0.19 &   11.47 &  22.14  &  1   & 3.20   & 3.20  & NED     &     K89    \\
PG\,1519+226 & 0.137 & $-$23.5 & $<$50 & 1.50  &$<$10.74 &  23.08  &      &$<$1.52 &       & ADDSCAN &     K89    \\
PG\,1552+085 & 0.119 & $-$23.3 & $<$50 & 0.80  &$<$10.62 &  22.69  &      &$<$1.80 &       & ADDSCAN &     K89    \\
PG\,1612+261 & 0.131 & $-$23.5 & 191  & 5.07   &   11.29 &  23.57  & 0.33 & 1.58   & 1.75  & FSC     &     K89    \\
PG\,1613+658 & 0.129 & $-$24.1 & 635  & 3.03   &   11.79 &  23.3   & 0.26 & 2.32   & 2.45  & S89     &     K89    \\
PG\,1617+175 & 0.114 & $-$23.6 & 102  & 1.09   &   10.89 &  23.04  &      & 1.97   &       & FSC     &     K89    \\
PG\,1626+554 & 0.133 & $-$23.3 & $<$50 & 0.17  &$<$10.72 &  22.11  &  1   &$<$2.47 &$<$2.47& ADDSCAN &     K89    \\
\hline
\end{tabular}
 \\
a. inferred from 20cm integrated emission minus flat 6cm core spectrum \\
b. IRAS Faint Source Catalog \\
c. NRAO VLA Sky Survey ({\tt http://www.cv.nrao.edu/nvss/}) \\
d. own high resolution VLA measurements \\
e. IRAS ADDSCAN analysis \\
f. NASA Extragalactic Database ({\tt http://nedwww.ipac.caltech.edu})
\end{table*}

\end{document}